\documentstyle[aps,prc,tighten,epsfig,floats]{revtex}

\begin{document}

\draft

\title{ Proton-deuteron elastic scattering at low energies }

\author{ C.~R.~Brune, W.~H.~Geist,\thanks{Present Address: Los
         Alamos National Laboratory, Los Alamos, NM 87545}
         H.~J.~Karwowski, E.~J.~Ludwig,
         K.~D.~Veal,$^*$ and M.~H.~Wood\thanks{Present Address:
         Thomas Jefferson National Accelerator Facility,
         Newport News, VA 23606} }
\address{ Department of Physics and Astronomy,
          University of North Carolina at Chapel Hill\\
          Chapel Hill, North Carolina 27599-3255, USA\\
          and Triangle Universities Nuclear Laboratory, 
          Durham, North Carolina 27708-0308, USA }
\author{ A.~Kievsky, S.~Rosati, and M.~Viviani}
\address{ Istituto Nazionale di Fisica Nucleare,
          Sezione di Pisa and Dipartimento di Pisa,
          Universita' di Pisa, I-56100 Pisa, Italy }

\date{\today}
\maketitle

\begin{abstract}
We present measurements of differential cross sections and
the analyzing powers $A_y$, $iT_{11}$, $T_{20}$, $T_{21}$, and
$T_{22}$ at $E_{c.m.}=431.3$~keV. In addition, an excitation
function of $iT_{11}(\theta_{c.m.}=87.8^\circ)$ for
$431.3\le E_{c.m.}\le 2000$~keV is presented.
These data are compared to calculations employing realistic
nucleon-nucleon interactions, both with and without three-nucleon forces.
Excellent agreement with the tensor analyzing powers and cross section
is found, while the $A_y$ and $iT_{11}$ data are found to be underpredicted
by the calculations.
\end{abstract}



\section{Introduction}
The three-nucleon continuum is an excellent
testing ground for nuclear interactions, since both precise
experiments and theoretical calculations are possible~\cite{Glo96}.
Comparisons between measured three-nucleon observables
and theoretical calculations provide stringent tests of
the underlying nucleon-nucleon (NN) and three-nucleon (3N) interactions.
Proton-deuteron elastic scattering at low energies offers
two particular advantages:
highly accurate experiments can be performed for a large number
of observables, and theoretical calculations are simplified
due to the energy being below the deuteron breakup threshold
and the relatively low number of partial waves required.

Past experimental studies have focussed on the differential cross section
and analyzing powers for p-d scattering at $E_{c.m.}=2$~MeV,
just below the deuteron breakup threshold~\cite{Knu93,Shi95}.
Theoretical calculations of these p-d scattering observables using
realistic NN and 3N interactions are now routinely possible~\cite{Kie94,Kie95}.
The calculations are found to reproduce the differential cross section
and tensor analyzing powers (TAPs) $T_{20}$, $T_{21}$, and $T_{22}$
quite well, while the vector analyzing powers (VAPs) $A_y$ and $iT_{11}$
are underpredicted by $\sim30\%$.
The available 3N forces are found to have little influence,
except for the $J^\pi={1\over 2}^+$ partial wave~\cite{Kie95,Kie96}.
The situation in neutron-deuteron scattering is
similar~\cite{Kie96,Mca94,Wit94},
although only the cross section and $A_y$ have been measured to date at
low energies. By far the most glaring difference between theory and
experiment is in the VAPs, a discrepancy which has come to be known
as the ``$A_y(\theta)$ puzzle''~\cite{Wit94}. This puzzle has been in
existence for about ten years now, and the resolution is still unclear.
It has been suggested that the discrepancy arises from inadequacies in
the NN potential in the $^3P_j$ waves~\cite{Wit91,Tor98}.
Other workers have considered the role of 3N force
effects~\cite{Kie96,Wit94,Bru98}.
So far, no conclusive solution to the problem has been found, although
most recent attention has been focussed on the role of the
3N interaction~\cite{Hub98,Kie99,Can00}.

It is desirable to determine these observables at
energies below $E_{c.m.}=2$~MeV,
as the influence of higher partial waves is strongly reduced, allowing the
dominant $S$- and $P$-waves to be investigated with greater clarity.
Also, when faced with disagreement between theory and experiment, it is
interesting to study the energy dependence of the discrepancy, since this
may provide clues as to its origin.
New data will place strong constraints on any proposed solution.
There is also considerable interest in obtaining low-energy data
for the determination of the p-d scattering lengths~\cite{Bla99}.

In this paper we present measurements of the differential cross
section and analyzing powers $A_y$, $iT_{11}$, $T{20}$, $T_{21}$ and $T_{22}$.
for p-d scattering at $E_{c.m.}=431.3$~keV.
These data are the first complete set of observables measured
at such a low energy;
previously only cross section~\cite{Bla99,Hut83} and
$A_y$ data~\cite{Hut83} have been obtained in this energy range.
We also present an energy excitation function of
$iT_{11}(\theta_{c.m.}=87.8^\circ)$ from $E_{c.m.}=431.3$~keV up to 2000~keV.
Some of the data presented here have appeared previously in
brief publications~\cite{Bru98,Kie97,Kar99}.

\section{Experimental procedures}
\label{sec:exp}

Measurements of the differential cross section
and the analyzing powers $A_y$, $iT_{11}$, $T_{20}$, $T_{21}$, and $T_{22}$
were carried out at the Triangle
Universities Nuclear Laboratory (TUNL).
The analyzing power $T_{22}$ was deduced from measurements
of $T_{20}$ and $A_{yy}$.
Achieving a center-of-mass energy of 431.3~keV requires either
a 650-keV proton beam incident on a deuteron target, or a 1300-keV
deuteron beam incident on a proton target (the small corrections
for energy loss in the target will be discussed later).
The beams, targets, and detection methods are discussed below.

\subsection{Low-energy proton beam}
\label{subsec:pbeam}

Beams of 650-keV protons were used for measurements of differential
cross sections and $A_y$. Polarized or unpolarized beams of
72-keV ${}^1{\rm H}^-$ ions were produced by the TUNL atomic beam polarized ion
source~\cite{Cle95}. These ions were then accelerated to 450~keV
by the minitandem accelerator~\cite{Bla93}.
The final energy of 650~keV for the now positive ions
was achieved by the $-200$-kV bias on
the 107-cm-diameter scattering chamber~\cite{Lud97} where the
measurements were performed.

Unpolarized proton beams were produced by bleeding hydrogen gas into
the ionizer of the ion source.
When polarized proton beams were used,
a Wien filter downstream from the ion source was utilized to
set the spin-quantization axis vertical in the laboratory.
Two spin states were used, with polarizations $p_Z\approx\pm 0.7$.
The desired hyperfine
states of atomic hydrogen were cycled approximately every second.
This technique minimizes the effects of slow changes in beam position,
target thickness, or amplifier gain on the measured analyzing powers.
Proton beam polarization was determined using the
${}^6{\rm Li}(\vec{p},{}^3{\rm He}){}^4{\rm He}$ reaction in
a polarimeter~\cite{Bru97} located at the rear of the scattering chamber.
The polarization was measured several times throughout the measurements
at an incident proton energy of 450~keV by lowering the
chamber bias voltage. The proton polarization was found to be constant within
$\pm 3\%$ throughout the measurements; the systematic error in the
proton polarization is estimated to be $\pm 4\%$.

The energy calibration of the proton beam
produced by the minitandem and high-voltage chamber system
has been established to $\pm 1$~keV, using the
240.0- and 340.5-keV resonances in
${}^{19}{\rm F}(p,\alpha\gamma)$ and the
405.4- and 445.8-keV resonances in
${}^{27}{\rm Al}(p,\gamma)$. The resonance energies were
taken from Ref.~\cite{Uhr85}.

\subsection{Deuteron beam}
\label{subsec:dbeam}

Polarized beams of 72-keV ${}^2{\rm H}^-$
ions were produced using the same
atomic-beam polarized ion source~\cite{Cle95} used for protons.
The beam was injected into the TUNL FN~tandem accelerator,
magnetically analyzed, and then delivered to a 62-cm diameter
scattering chamber.

The same Wien filter mentioned previously was used to
control the spin-quantization axis on target.
The spin axis was longitudinal for the $T_{20}$ measurements,
normal to the reaction plane for $iT_{11}$ and $A_{yy}$, and
$45^\circ$ offset from longitudinal in the reaction plane for $T_{21}$.
Three deuteron beam polarization states were produced with the atomic beam
polarized ion source: a maximum positive, a maximum
negative, and an unpolarized state.
The TAP data were obtained with $p_{ZZ}\approx\pm 0.70$ and
$p_Z\approx\mp 0.25$, while beams with $p_{Z}\approx\pm 0.55$ and
$|p_{ZZ}|\le 0.05$ were used for the $iT_{11}$ measurements.
The spin states were also cycled approximately once every second.
The beam polarization for the TAP data was determined using the
${}^3{\rm He}(\vec{d},p)$ reaction in an online polarimeter located behind the
scattering chamber~\cite{Ton80}.
Deuteron beam vector polarization was determined online via the
${}^{12}{\rm C}(\vec{d},p)$ reaction in a different polarimeter
located behind the scattering chamber.
The effective $iT_{11}$ for this reaction at $E_d=1.3$~MeV
has been calibrated relative to the ${}^3{\rm He}(\vec{d},p)$
polarimeter at $E_d=12$~MeV.
More information on the use of the
${}^{12}{\rm C}(\vec{d},p)$ reaction for deuteron
vector polarimetry is available in Ref.~\cite{Woo00}.
For the $iT_{11}(87.8^\circ)$ measurements the polarization was
very stable over time, and was monitored
periodically with the ${}^3{\rm He}(\vec{d},p)$
polarimeter at $E_d=12$~MeV.
The absolute uncertainty in all of the deuteron beam polarizations is
estimated to $\pm 3\%$.
The unpolarized deuteron and proton beams used for the
${}^1{\rm H}(d,d)$ cross section measurements were produced by
a direct extraction negative ion source.

The energy calibration
of the beams produced by the FN~tandem is determined within $\pm0.1\%$
from the magnetic
field in the $52^\circ$ bending magnet, measured by an NMR magnetometer.
The magnetometer was calibrated using the ${}^7{\rm Li}(p,n)$
threshold ($E_p=1880.443\pm 0.020$~keV~\cite{Whi85}) and the
${}^{19}{\rm F}(p,n)$ threshold ($E_p=4234.3\pm 0.8$~keV~\cite{Mar66}).

\subsection{Targets}

The measurements were performed by bombarding thin hydrogenated or deuterated
carbon foils. The targets used for analyzing power measurements
consisted of approximately
$1\times 10^{18}$ and $1.5\times 10^{18}$ hydrogen isotope
and carbon atoms/cm$^2$, respectively.
The targets used for the relative and absolute
cross section measurements consisted of
approximately $5\times 10^{17}$ and $8\times 10^{17}$ hydrogen isotope
and carbon atoms/cm$^2$, respectively.
In order to calculate the energy loss in the targets it is necessary
to know their carbon and hydrogen isotope thickness and the stopping power.
The carbon content was determined by comparison with carbon foils of
known thickness while the hydrogen thickness was determined either
from the d-p elastic scattering cross section, or from the p-p
elastic scattering cross section.
For the analyzing power measurements
the beam-energy loss in the targets is $\approx 10$~keV,
while for the cross section measurements, the beam-energy
loss was $\approx 5$~keV.
In all cases the incident energy was adjusted so that the
mean energy in the center of the target corresponded
to $E_{c.m.}=431.3\pm 0.8$~keV.
Here the quoted error includes the uncertainties in the incident energy
and also the energy loss in the target.
It should be noted that the use of thin targets is very important at
low energies for minimizing energy loss and straggling effects.
The differential cross section is particularly sensitive to the energy,
a point which will be discussed quantitatively later in this paper.

\subsection{Detection methods}

The reaction products were detected using 100-$\mu$m-thick Si surface
barrier detectors.
Left-right symmetric detector configurations were utilized for
all of the measurements.
Count rates in the detectors were controlled by
varying the solid angle (0.2-5~msr) and beam current (10-150~nA).
In some cases, thin mylar foils (2-5~$\mu$m) were placed over the detectors
to either (1) increase the energy separation between proton and deuteron
peaks or (2) stop heavy recoils resulting from
${}^{12}{\rm C}(d,{}^{12}{\rm C})d$, ${}^{12}{\rm C}(d,{}^{13}{\rm C})p$, or
${}^{12}{\rm C}(d,{}^{13}{\rm N})n$ reactions.
Measurements at far forward angles in the laboratory are limited
by the high flux of elastic-scattered particles from carbon.
For the proton-beam measurements in the high-voltage scattering
chamber, detector signals were sent to ground potential via
fiber-optic connections.

\begin{figure}
\begin{center}
\includegraphics[bbllx=43,bblly=247,bburx=433,bbury=461,%
width=3.4in]{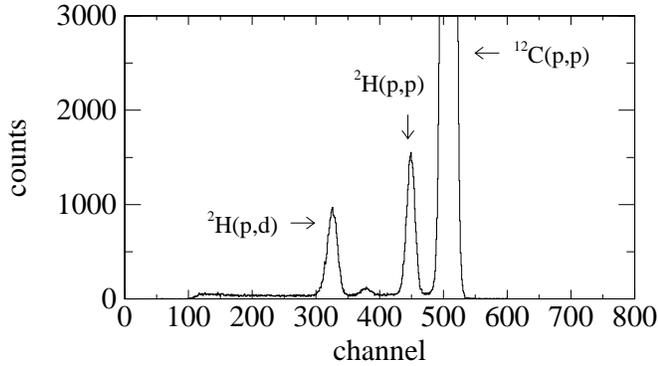}
\end{center}
\caption{Charged-particle spectrum obtained
at $\theta_{lab}=30^\circ$
with 650-keV protons incident on a carbon-deuterium target.
The small peak near channel 380 results from ${}^1{\rm H}(p,p)$.}
\label{fig:spec_pond}
\end{figure}

\begin{figure}
\begin{center}
\includegraphics[bbllx=43,bblly=247,bburx=433,bbury=461,%
width=3.4in]{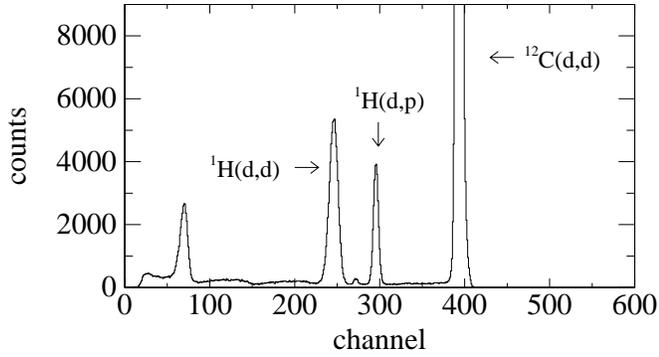}
\end{center}
\caption{Charged-particle spectrum obtained
at $\theta_{lab}=25^\circ$
with 1300-keV deuterons incident on a carbon-hydrogen target.
The peak near channel 70 also results from
${}^1{\rm H}(d,d)$, but the lower-energy deuteron peaks
were not analyzed due to the larger backgrounds.
The small peak near channel 270 results from
${}^{12}{\rm C}(d,p){}^{13}{\rm C}(\mbox{3.09~MeV})$.
The broad structures below channel 250 result from heavy-ion
recoils from the target.}
\label{fig:spec_donp1}
\end{figure}

\begin{figure}
\begin{center}
\includegraphics[bbllx=43,bblly=247,bburx=433,bbury=461,%
width=3.4in]{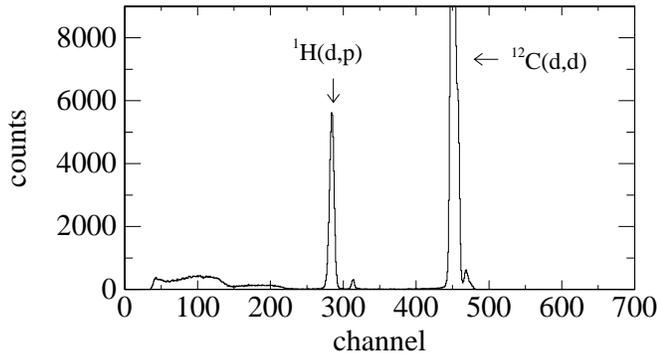}
\end{center}
\caption{Charged-particle spectrum obtained
at $\theta_{lab}=35^\circ$
with 1300-keV deuterons incident on a carbon-hydrogen target.
The small peak near channel 310 results from
${}^{12}{\rm C}(d,p){}^{13}{\rm C}(\mbox{3.09~MeV})$.
The broad structures below channel 250 results from heavy-ion
recoils from the target.}
\label{fig:spec_donp2}
\end{figure}

For the cross section, $T_{20}$, $T_{21}$, and $A_{yy}$ measurements
the elastic scattering yields were determined from the
energy spectra.
Sample energy spectra are shown in
Figs.~\ref{fig:spec_pond}-\ref{fig:spec_donp2}.
It should be noted that the low-energy heavy recoils seen in
Figs.~\ref{fig:spec_donp1} and~\ref{fig:spec_donp2}
can be eliminated if necessary by using the
aforementioned mylar foils.
The primary background under the p-d elastic scattering peaks
at all angles is the low-energy tail from carbon elastic scattering.
Pulsers were inserted into the detector electronics to facilitate dead-time
corrections, which were typically less than 5\% but always
less than 10\%.
Peak yields were extracted using linear least-squares fits to the
background on either side of the peak.
For the analyzing power measurements, the background subtraction and
dead-time corrections were performed separately for each spin state.

\begin{figure}
\begin{center}
\includegraphics[bbllx=54,bblly=247,bburx=424,bbury=466,%
width=3.4in]{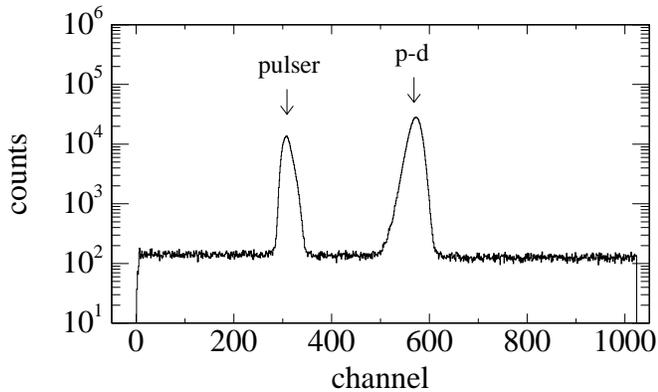}
\end{center}
\caption{Proton-deuteron time-of-flight spectrum obtained
for $\theta_{c.m.}=87.8^\circ$
with 1300-keV deuterons incident on a carbon-hydrogen target.
The horizontal axis calibration is approximately 0.4~ns/channel.}
\label{fig:spec_tac}
\end{figure}

A different technique was used for the VAPs $A_y$ and $iT_{11}$ since
the magnitudes of these observables are roughly a factor of 10
smaller than for the TAPs.
The scattered deuterons and protons were detected in coincidence using
two pairs of silicon surface barrier detectors placed at symmetric
angles on either side of the incident beam.
The angles of the detectors were set
to observe either protons or deuterons in the
more forward detectors in coincidence with deuterons or protons
detected in the more backward detector on the opposite side of the beam.
A major benefit of this method is that data can be acquired at a much higher
rate because the events resulting from carbon elastic scattering are
essentially completely eliminated. Note that a factor of 10 reduction in
magnitude of the analyzing power requires a factor of 100 increase in the
number of counts if the same relative statistical accuracy is desired.
An additional advantage of the technique is that the background under
the peak of interest is much reduced, lowering the possibility of false
asymmetries resulting from the asymmetry in the background.
Histograms of the time difference between the fast timing signals
from each coincident pair of detectors were stored for each spin state.
The time resolution for the coincident proton-deuteron peaks
was $\approx 8$~ns, with backgrounds $<3\%$.
A sample time spectrum is shown in Fig.~\ref{fig:spec_tac}.
Dead-time corrections ($<3\%$) were determined by
sending test pulses to the detector preamplifiers with time delays
adjusted to give distinct peaks in the time spectra.
Peak yields for each spin state were extracted using linear least-squares
fits to the background on either side of the peak, and were
corrected for dead time.
The possible influence of channel (time) dependent dead-time corrections
was calculated and found to be negligible.

\subsection{Measurements}

The incident beam energy was adjusted to yield $E_{c.m.}=431.3$~keV
in center of the target, except for the $iT_{11}(87.8^\circ)$ measurements
where the incident deuteron energy varied
over $1.3\le E_d \le 6.0$~MeV.
The beams were collimated to produce a 
2~mm~(horizontal)~$\times$~4~mm~(vertical)
beamspot on the center of the target.
The $0^\circ$ position for each detector was determined by measuring the
elastic scattering from carbon on either side of the beam for angles near
$0^\circ$. The systematic uncertainty in the angular positioning
is estimated to be $\pm 0.1^\circ$.
The number of incident particles was determined by
beam current integration.
For the analyzing power data and relative cross section data 
the targets were replaced
approximately every 12~hours, or when the hydrogen isotope content
decreased by $\approx 30$\%.

Relative cross sections were determined with a proton beam in the high-voltage
scattering chamber. The measurements were performed using a movable
pair of left-right symmetric detectors, and a fixed pair located
out of the plane of the other detectors at $\theta_{lab}=37.5^\circ$.
The yield of protons and deuterons from p-d elastic scattering in
the fixed pair was used to normalize the yields in the movable pair.

Absolute p-d cross sections were determined relative to the proton-proton
elastic scattering cross section at a nominal proton beam
energy of 2600~keV.
Detector pairs were placed at $\theta_{lab}=25^\circ$ and $35^\circ$
where both cross sections vary smoothly with angle.
These angles give d-p elastic scattering measurements at
$\theta_{c.m.}=82.6^\circ$, $110^\circ$, and $130^\circ$.
A direct extraction negative ion source was
used to inject beam into the tandem accelerator for these measurements.
The beam could be switched between
1300-keV deuterons to 2600-keV protons in about one minute by only
adjusting two parameters:
(1) the magnetic field in the bending magnet between the source and
the accelerator, and (2) changing the accelerator terminal voltage.
Note that the magnetic field in the $52^\circ$ analyzing magnet,
which defines the beam energy, is unchanged throughout this procedure.
The deuteron and proton beams were switched several times for each target,
using an integrated charge of $\approx 1$~$\mu$C per measurement.
For these measurements accurate relative beam current integration is essential.
Current was integrated from the target rod (biased to $+100$~V)
and a plate at the rear of the chamber (suppressed by a shroud at $-100$~V).
As a test of the technique, we also measured using the same procedure
${}^{197}{\rm Au}(d,d)$ and ${}^{197}{\rm Au}(p,p)$ elastic scattering
at $\theta_{\rm lab}=140^\circ$, where the cross sections are described
to a very high accuracy by the Rutherford formula.
A target consisting of $2\times 10^{17}$ Au atoms/cm$^2$ was utilized
for these measurements.
The results agreed within 0.8\% with the ratio expected for Rutherford
scattering ($\approx 1\%$ corrections due to electron
screening~\cite{Hut85} were taken into account).

\section{Data Analysis and Results}
\label{sec:res}

\subsection{Cross section measurements}
\label{subsec:wexp}

\begin{figure}
\begin{center}
\includegraphics[bbllx=127,bblly=96,bburx=494,bbury=601,angle=270,
width=3.4in]{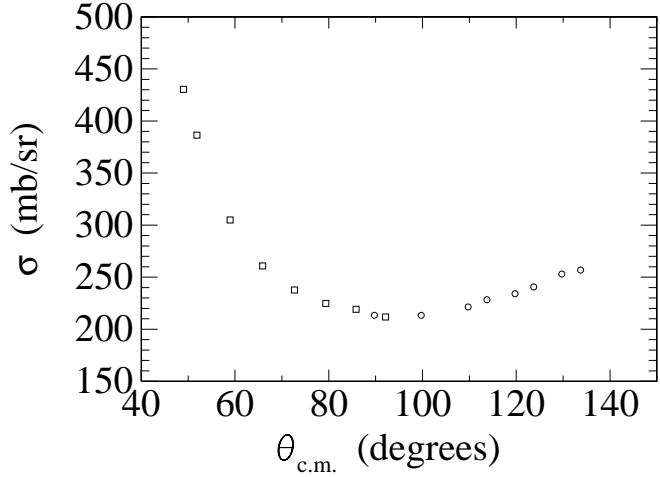}
\end{center}
\caption{Relative proton-deuteron differential cross section data
obtained $E_{c.m.}=431.3$~keV. The squares and circles result from
detected protons and deuterons, respectively.
These data have been normalized to the absolute cross section measurements.}
\label{fig:rel_sig}
\end{figure}

The relative cross section data as a function of center-of-mass angle
are shown in Fig.~\ref{fig:rel_sig}.
These data have been normalized to the absolute cross section measurements.

Absolute p-d cross sections were determined at $\theta_{lab}=25^\circ$ and
$35^\circ$  relative to the p-p scattering measurements
at a nominal proton beam energy of 2600~keV.
The p-p cross section for these energies and angles appears to
be well understood. Our procedure is to utilize the Nijmegen
energy-dependent partial-wave analysis~\cite{Sto93,Ren97}
to generate cross sections for our energy and angles.
A thorough discussion of the data upon which this analysis is based is
given in Ref.~\cite{Ber88}.
The p-p cross section data for $E_p<10$~MeV which are included in
the Nijmegen analysis generally have absolute uncertainties $<0.5\%$
and are reproduced by the fit within this error.
In addition there are the p-p cross section data of
Knecht, Dahl, and Messelt~\cite{Kne66} which include measurements
at $E_p=2425$~keV, very close to the energy of the present experiment.
The quoted experimental uncertainties vary in the range 0.1-0.3\%.
Although not used in the Nijmegen analysis, the Nijmegen energy-dependent
fit agrees with the results of Ref.~\cite{Kne66} at 2425~keV within 0.4\% for
the angles of our interest.
The systematic uncertainty in the absolute p-d cross sections is
estimated to be $\pm 1.1\%$;
the contributions to this uncertainty are summarized
in Table~\ref{tab:sig_errors}.
The assumed p-p cross sections and the p-d cross sections derived
from them are given in Table~\ref{tab:abs_sig}.

\subsection{Analyzing power measurements}
\label{subsec:ayexp}

Analyzing powers were determined from the particle yields measured for
each spin state. The yields for each spin state were corrected for
background and dead time, and were normalized
by the number of incident particles.
Tests for false asymmetries using tensor-polarized deuteron beams
were carried out by measuring
${}^{197}$Au(d,d) scattering at $\theta_{lab}=40^\circ$ under identical
conditions as the p-d TAP measurements.
For $E_d=1.3$~MeV, all of analyzing powers for ${}^{197}$Au(d,d)
are expected to be $<10^{-4}$~\cite{Kam85}.
The results were consistent with zero at the level
of $5\times 10^{-4}$, the statistical uncertainty of the measurement.

\begin{figure}
\begin{center}
\includegraphics[bbllx=19,bblly=244,bburx=430,bbury=461,angle=0,
width=3.4in]{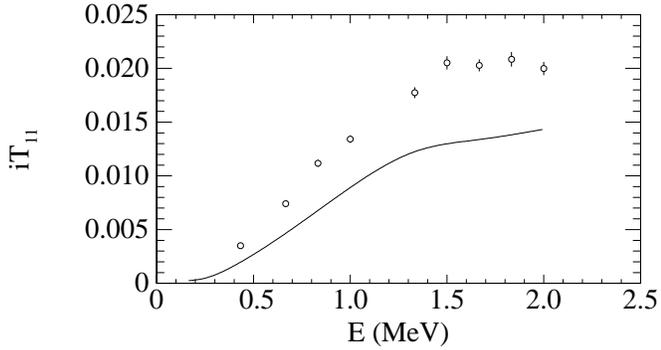}
\end{center}
\caption{The measured $iT_{11}$ (circles) for $\theta_{c.m.}=87.8^\circ$
plotted versus center-of-mass energy.
The solid curve is a theoretical calculation using the AV18+UR-IX potential.}
\label{fig:excit}
\end{figure}

For each pair of left-right symmetric detectors,
the analyzing powers were determined as follows.
We first define the following ratios
\begin{eqnarray}
L &=& Y^{(1)}_L/Y^{(2)}_L \mbox{~and} \\
R &=& Y^{(1)}_R/Y^{(2)}_R,
\end{eqnarray}
where $Y^{(i)}_L$ and $Y^{(i)}_R$ are the particle yields in the left
and right detector, respectively, while $i=1,2$ denotes the spin state.
Then
\begin{eqnarray}
T_{20} &=& {1\over\sqrt{2}}\left[{L+R-2\over p_{ZZ}^{(1)}-
  {1\over 2}(L+R)p_{ZZ}^{(2)}}\right], \\
T_{21} &=& {1\over\sqrt{3}}\left[{1-L\over p_{ZZ}^{(1)}-Lp_{ZZ}^{(2)}}
  -{1-R\over p_{ZZ}^{(1)}-Rp_{ZZ}^{(2)}}\right], \\
A_{yy} &=& {L+R-2\over p_{ZZ}^{(1)}-{1\over 2}(L+R)p_{ZZ}^{(2)}},
  \label{eq:ayy}\\
iT_{11} &=& {1\over\sqrt{3}}\left[{L-R\over 2p_Z^{(1)}-(L+R)p_Z^{(2)}}\right],
  \mbox{~and} \label{eq:it11}\\
A_y &=& {2(L-1)(R-1)\over(p_Z^{(2)}-p_Z^{(1)})(L-R)},
\end{eqnarray}
where $p_Z^{(i)}$ and $p_{ZZ}^{(i)}$ are the vector and tensor
polarizations for spin state $i$. It should be noted that
Eqs.~(\ref{eq:ayy}) and~(\ref{eq:it11}) are approximate; for the
analyzing powers and polarizations encountered in this experiment
the corrections are negligible.
Values for the TAP $T_{22}$ were derived from $A_{yy}$ and $T_{20}$
using $T_{22}=-A_{yy}/\sqrt{3}-T_{20}/\sqrt{6}$.
The analyzing power data are shown in Figs.~\ref{fig:tensor}
and~\ref{fig:vector}.
The measured $iT_{11}(87.8^\circ)$ excitation function is
shown in Fig.~\ref{fig:excit}.

\section{Discussion}
\label{sec:dis}

\begin{figure}
\begin{center}
\includegraphics[bbllx=52,bblly=89,bburx=347,bbury=571,angle=0,
width=3.4in]{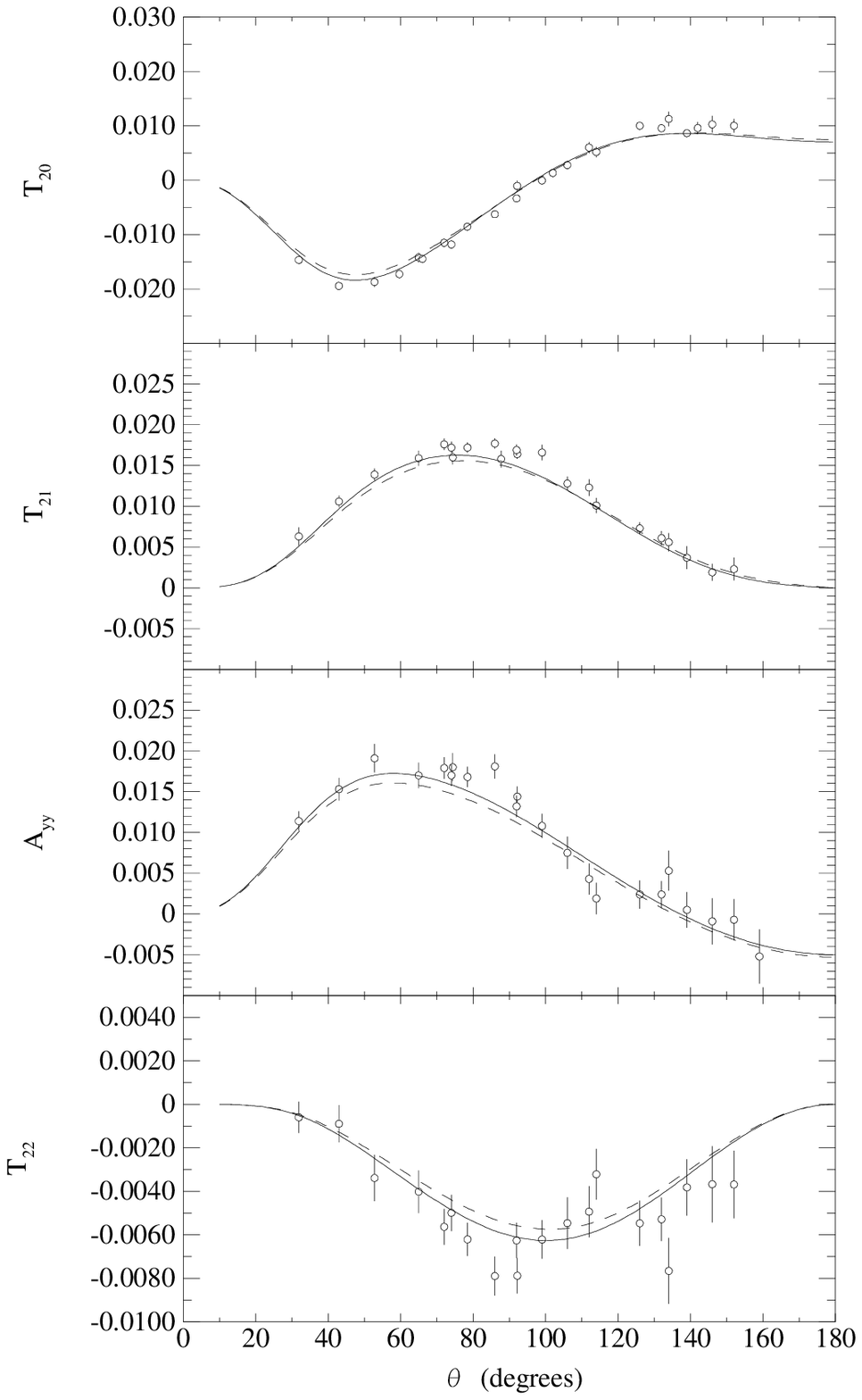}
\end{center}
\caption{The measured TAP data (circles) and
theoretical calculations using the AV18 (dashed line)
and AV18+UR-IX (solid line) potentials.}
\label{fig:tensor}
\end{figure}

\begin{figure}
\begin{center}
\includegraphics[bbllx=29,bblly=98,bburx=476,bbury=603,angle=0,
width=3.4in]{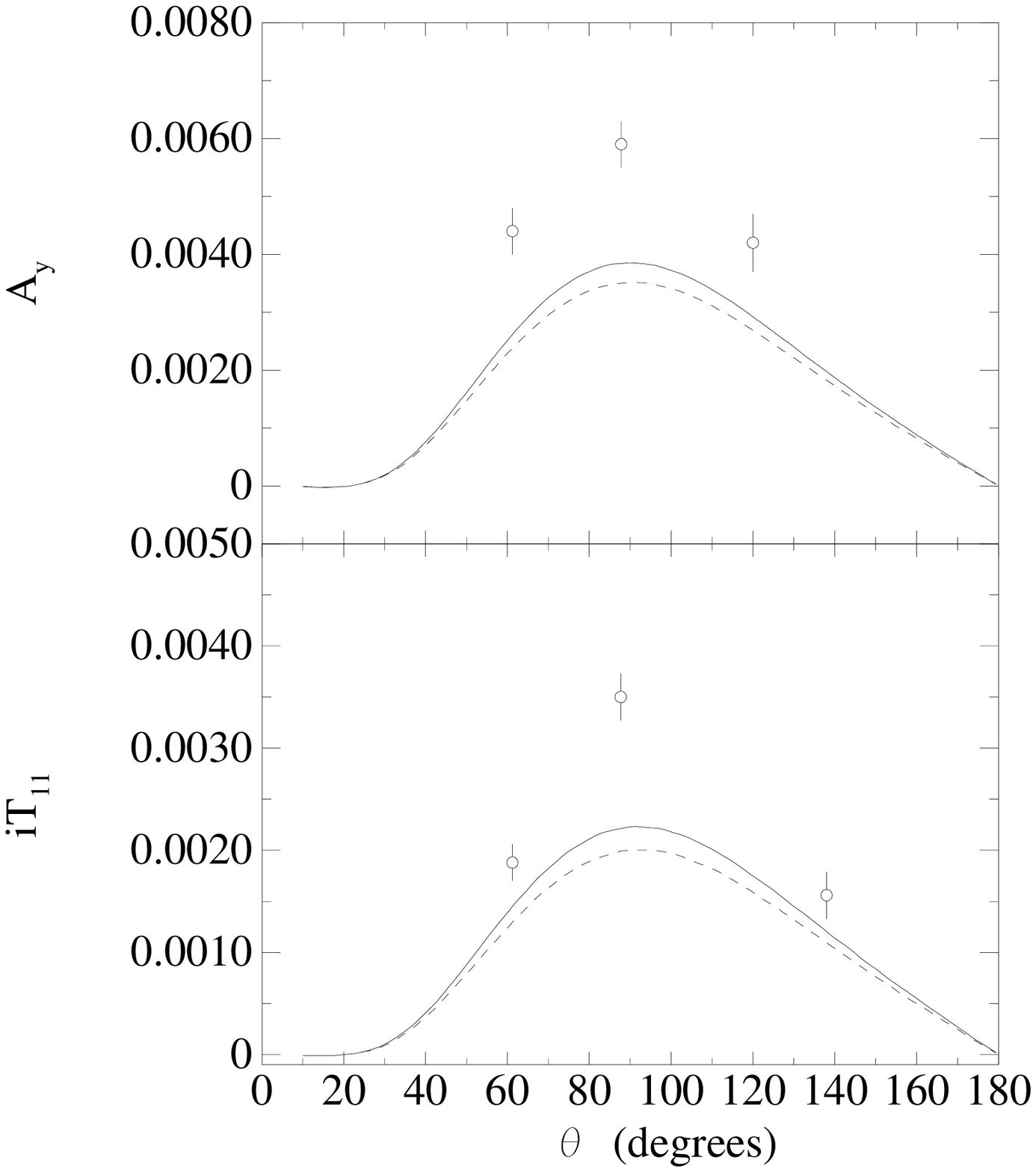}
\end{center}
\caption{The measured VAP data (circles) and
theoretical calculations using the AV18 (dashed line)
and AV18+UR-IX (solid line) potentials.}
\label{fig:vector}
\end{figure}

\begin{figure}
\begin{center}
\includegraphics[bbllx=30,bblly=241,bburx=433,bbury=461,angle=0,
width=3.4in]{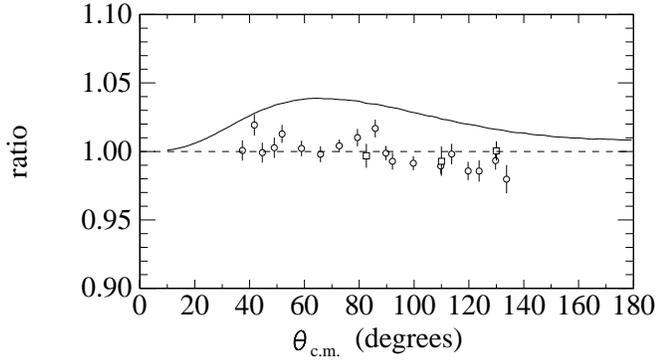}
\end{center}
\caption{Ratios of cross sections to the theoretical cross sections
calculated using the AV18+UR-IX potential. The experimental cross sections
are shown as circles and squares for the relative and absolute
measurements, respectively. The ratio of the AV18 cross section
to the AV18+UR-IX cross section is shown by the solid line.}
\label{fig:ratio}
\end{figure}

These experimental results are compared to calculations
utilizing the Pair-Correlated Hyperspherical Harmonic
basis~\cite{Kie93} to construct the scattering wave function,
and the Kohn variational principle to determine
the scattering matrix elements~\cite{Kie94}.
The calculations were performed using the AV18 NN potential~\cite{Wir95}
and with AV18 plus the 3N interaction
of Urbana (UR-IX)~\cite{Pud95}, and are shown in
Figs.~\ref{fig:tensor},~\ref{fig:vector}, and~\ref{fig:ratio}.
All of the cross sections in Fig.~\ref{fig:ratio} have been divided
by the cross section calculated using the AV18+UR-IX potential, so
that the small differences are more apparent. The cross section
measurements are seen to be in excellent agreement with the calculations
using the AV18+UR-IX potential. The calculation using the AV18 potential
alone differs from the measurements by 3-4\%,
a small but statistically-significant amount.
As shown in Fig.~\ref{fig:tensor}, the TAP data are in reasonably good
agreement with calculations using either AV18 or AV18+UR-IX.
It is, however, significant to note that the AV18+UR-IX potential
gives slightly better agreement for all of the TAPs.

The VAP results are shown Fig.~\ref{fig:vector}. As we have previously
reported~\cite{Bru98}, the calculations using both the AV18 and AV18+UR-IX
potentials underpredict the data by $\approx 40$\%.
This discrepancy has now been observed for a wide range of energies
for both $n-d$ and $p-d$ scattering. The origin of this
``$A_y$ puzzle'' is not clear.
Specific efforts to study this problem have been undertaken recently
by employing new forms of 3N forces~\cite{Kie99,Can00}.
Moreover in Ref.~\cite{Epe00} a new NN interaction constructed from chiral
perturbation theory has been shown to give a better description 
of the VAPs at low energy.

The measured energy dependence of $iT_{11}(87.8^\circ)$,
shown in Fig.~\ref{fig:excit}, will have to be explained by any
proposed solution to the $A_y$ puzzle.
It is seen that the AV18+UR-IX potential consistently
underpredicts $iT_{11}$ for all energies below the deuteron breakup
threshold. Although not shown, the AV18 potential calculation
lies $\approx10$\% lower than for AV18+UR-IX at these energies as well.
We also note the the present measurements at $E_{c.m.}=1.67$~ and 2.00~MeV
are in excellent agreement with previous data~\cite{Knu93,Shi95}.

\begin{figure}
\begin{center}
\includegraphics[bbllx=128,bblly=90,bburx=490,bbury=622,angle=270,
width=3.4in]{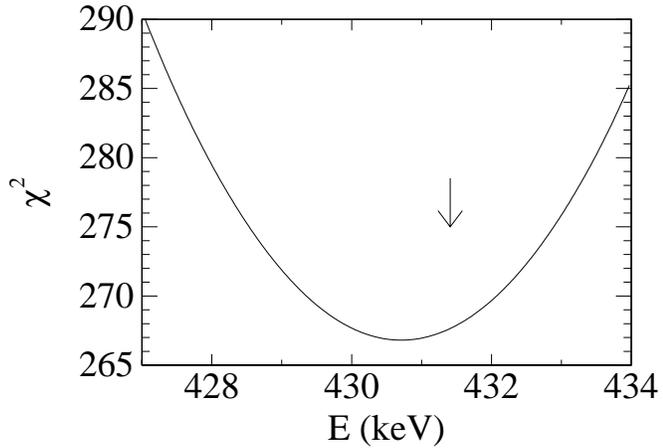}
\end{center}
\caption{Plot of $\chi^2$ versus center-of-mass energy
for the comparison of the experimental
data to theoretical calculations at different energies.
The arrow indicates the actual energy of the experimental data.}
\label{fig:chi2_e}
\end{figure}

\begin{figure}
\begin{center}
\includegraphics[bbllx=128,bblly=90,bburx=485,bbury=625,angle=270,
width=3.4in]{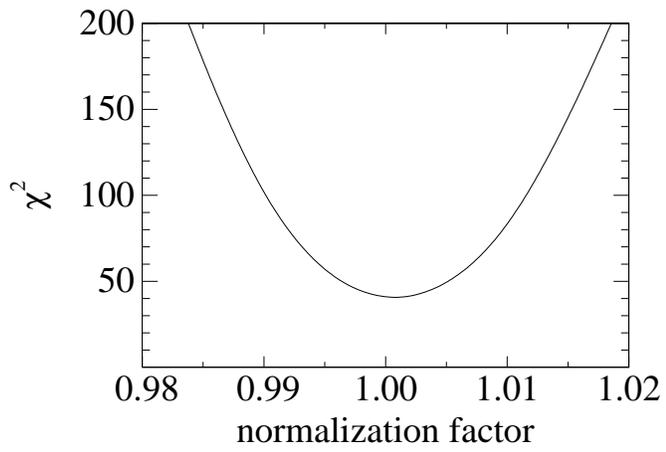}
\end{center}
\caption{Plot of $\chi^2$ versus the cross section normalization factor.}
\label{fig:chi2_s}
\end{figure}

The comparison between the present data and theoretical calculations
at $E_{c.m.}=431.3$~keV can be made more quantitative through the use of the
$\chi^2$ parameter. For the $N=94$ data points
the AV18 calculation yields $\chi^2=844$, while
the AV18+UR-IX calculation yields $\chi^2=268$.
The result that $\chi^2/N$ is much greater than one even
for the AV18+UR-IX calculation is expected since there are large
discrepancies for the VAPs as well as smaller but
statistically-significant differences in the other observables.
A more detailed discussion of $\chi^2$ comparisons and also phase-shift
analysis by $\chi^2$ minimization will be the subject of another
publication. Here we comment on two systematic uncertainties in the
experimental data and how they affect the calculated $\chi^2$.
Considerable attention was given to the determination of the energy
of the experiment, $E_{c.m.}=431.3\pm 0.8$~keV. In Fig.~\ref{fig:chi2_e}
we show the sizable effect on $\chi^2$ of varying the energy of the theoretical
calculations with the AV18+UR-IX interaction.
It is clearly important for the experimental energy to be determined
as precisely as possible. We have also investigated the importance of
the absolute normalization of the experimental cross sections.
In Fig.~\ref{fig:chi2_s} we show the effect on $\chi^2$ of varying
the normalization of the cross section.
Here $\chi^2$ is calculated for the 22 cross section points.
Again very significant effects
are observed.

\section{Conclusions}
\label{sec:conc}

We have measured differential cross sections and
the analyzing powers $A_y$, $iT_{11}$, $T_{20}$, $T_{21}$, and
$T_{22}$ at $E_{c.m.}=431.3$~keV, as well as an excitation
function of $iT_{11}(\theta_{c.m.}=87.8^\circ)$ for
$431.3\le E_{c.m.}\le 2000$~keV. The 431.3-keV data
comprise the lowest-energy complete set of cross section and analyzing power
observables for nucleon-deuteron scattering.
The cross section and TAPs are
found to be in good agreement with calculations using the
AV18 NN  interaction and the Urbana-IX 3N interaction.
The VAPs $A_y$ and $iT_{11}$ are found to be underpredicted by
$\approx 40$\%. At this time it is not clear whether these discrepancies
result from inadequacies in the assumed NN interaction, the 3N
interaction, or some other source.
In this context, the very low energy data presented here represent
a very stringent constraint for any new theoretical model of the
nuclear interaction.

\section*{Acknowledgments}
\label{sec:ack}

The authors would like to thank B.~J.~Crowe and B.~M.~Fisher for
their assistance in the data collection process,
and M.C.M. Rentmeester for supplying the Nijmegen p-p scattering predictions.
This work was supported in part by the U.S. Department of 
Energy, Office of High Energy and Nuclear Physics, under 
grant No. DE-FG02-97ER41041.

\tighten

\narrowtext

\begin{table}
\caption{Systematic errors in the absolute cross section
determination.}
\begin{tabular}{lc}
source                   & error \\ \hline
p-p cross section        & 0.5\% \\
target uniformity        & 0.5\% \\
beam-current integration & 0.8\% \\ \hline
total                    & 1.1\%
\end{tabular}
\label{tab:sig_errors}
\end{table}

\begin{table}
\caption{Laboratory proton-deuteron cross sections with statistical
errors determined at $E_d=1294$~keV together with the
proton-proton elastic scattering cross sections assumed for $E_p=2589$~keV.}
\begin{tabular}{lccc}
& p(p,p)  & p(d,d)  & p(d,p)  \\
& (mb/sr) & (mb/sr) & (mb/sr) \\ \hline
$\sigma (\theta_{lab}=25.0^\circ)$ & 518.9 & $2240\pm 20$ & $926\pm 10$ \\
$\sigma (\theta_{lab}=35.0^\circ)$ & 493.4 & -            & $729\pm 8$
\end{tabular}
\label{tab:abs_sig}
\end{table}

\end{document}